\documentclass[twocolumn]{aastex631}

\usepackage{xcolor}
\usepackage[normalem]{ulem}
\usepackage{comment}
\usepackage{amsmath}

\received{December 22, 2023}
\accepted{January 23, 2024}

\shorttitle{Discovery of a spatially resolved disk wind}
\shortauthors{Bajaj et al.}

\graphicspath{{./}{figures/}}

\begin{document}

\title{JWST MIRI-MRS Observations of T\,Cha: Discovery of a Spatially Resolved Disk Wind}


\correspondingauthor{Naman S. Bajaj}
\email{namanbajaj@arizona.edu}

\author[0000-0003-3401-1704]{Naman S. Bajaj}
\affiliation{Lunar and Planetary Laboratory, The University of Arizona, Tucson, AZ 85721, US}

\author[0000-0001-7962-1683]{Ilaria Pascucci}
\affiliation{Lunar and Planetary Laboratory, The University of Arizona, Tucson, AZ 85721, US}

\author{Uma Gorti}
\affiliation{NASA Ames Research Center, Moffett field, CA 94035, US}
\affiliation{Carl Sagan Center, SETI Institute, Mountain View, CA 94043, US}

\author[0000-0001-6410-2899]{Richard Alexander}
\affiliation{School of Physics and Astronomy, University of Leicester, University Road, Leicester LEI 7RH, UK}

\author[0000-0003-0330-1506]{Andrew Sellek}
\affiliation{Institute of Astronomy, Madingley Road, Cambridge CB3 0HA, UK}
\affiliation{Leiden Observatory, Leiden University, 2300 RA Leiden, The Netherlands}

\author{Jane Morrison}
\affiliation{Steward Observatory, The University of Arizona, Tucson, AZ 85721, US}

\author[0000-0001-8612-3236]{Andras Gaspar}
\affiliation{Steward Observatory, The University of Arizona, Tucson, AZ 85721, US}

\author{Cathie Clarke}
\affiliation{Institute of Astronomy, Madingley Road, Cambridge CB3 0HA, UK}

\author[0000-0001-8184-5547]{Chengyan Xie}
\affiliation{Lunar and Planetary Laboratory, The University of Arizona, Tucson, AZ 85721, US}

\author[0000-0002-4687-2133]{Giulia Ballabio}
\affiliation{Astronomy Unit, School of Physics and Astronomy, Queen Mary University of London, London E1 4NS, UK}
\affiliation{Astrophysics Group, Imperial College London, Blackett Laboratory, Prince Consort Road, London SW7 2AZ, UK}

\author[0000-0003-0777-7392]{Dingshan Deng}
\affiliation{Lunar and Planetary Laboratory, The University of Arizona, Tucson, AZ 85721, US}

\begin{abstract}
Understanding when and how circumstellar disks disperse is crucial to constrain planet formation and migration. Thermal winds powered by high-energy stellar photons have long been theorized to drive disk dispersal. However, evidence for these winds is currently based only on small ($\sim$3-6 km~s$^{-1}$) blue-shifts in [Ne\,II] 12.81 \micron{} lines, which does not exclude MHD winds. We report JWST MIRI MRS spectro-imaging of T\,Cha, a disk with a large dust gap ($\sim$30 au in radius) and blue-shifted [Ne\,II] emission. We detect four forbidden noble gas lines, [Ar\,II], [Ar\,III], [Ne\,II], and [Ne\,III], of which [Ar\,III] is the first detection in any protoplanetary disk. We use line flux ratios to constrain the energy of the ionizing photons and find that Argon is ionized by EUV whereas Neon is most likely ionized by X-rays. After performing continuum and Point Spread Function (PSF) subtraction on the IFU cube, we discover a spatial extension in the [Ne\,II] emission off the disk continuum emission. This is the first spatially resolved [Ne\,II] disk wind emission. The mostly ionic spectrum of T\,Cha, in combination with the extended [Ne\,II] emission, points to an evolved stage for any inner MHD wind and is consistent with the existence of an outer thermal wind ionized and driven by high-energy stellar photons. This work acts as a pathfinder for future observations aiming at investigating disk dispersal using JWST.
\end{abstract}


\section{Introduction} \label{sec:intro}
According to the classic picture of protoplanetary disk evolution, disk material is depleted through viscous accretion onto the star \citep[e.g.][]{Hartmann2016} and via thermal winds \citep[also known as photoevaporative winds, e.g. ][]{Alexander2014}. Viscous accretion is a process in which the gas viscosity (or turbulent stresses) creates friction between different layers of material, causing a shearing motion. The energy of the shearing motion is dissipated as heat, and the gas falls deeper in the potential well, causing it to accrete onto the star \citep[e.g.,][]{Pringle1981,Ruden1991}. On the other hand, photoevaporative (PE) winds are gaseous outflows driven away from the disk due to high-energy ($>$6 eV) radiation from the star. These winds are only expected to originate outward of $\sim$0.1-0.2 r$_G$, where r$_G$ is the gravitational radius where the thermal energy becomes large enough to overcome the gravitational potential of the star \citep[e.g.,][]{Gorti2016}. When the accretion rate falls below the PE wind mass loss rate, the gas supply to the inner disk is limited, and a gap is opened, after which the disk is quickly cleared from inside-out \citep[e.g.,][]{Clarke2001}. In this scenario, there is an intermediate short-lived stage of a protoplanetary disk where there is a large gap in the gas and dust \citep[e.g.,][for a discussion]{Ercolano2017}. It has also been shown through several studies that the interaction of planets with the disk can excite a spiral structure, which can then generate shocks and open a gap \citep[][for a recent review]{Bae2023}. However, regardless of whether these gaps are formed by photoevaporation or planets, constraining the PE wind mass loss rate will inform us of the lifetime of the gas disk, which in turn governs the time left for giant planet formation and migration \citep[e.g.,][]{Testi2014}.

More recently, disk simulations with detailed microphysics found that accretion is mainly driven by magnetohydrodynamic (MHD) disk winds \citep[e.g.,][for recent reviews]{Lesur2021,Lesur2023,Pascucci2023}. These MHD winds are magnetically induced gas flows that remove significant mass and angular momentum from the disk, leading to accretion onto the star. Along with theoretical advances, several disk wind diagnostics have been identified \citep[e.g.,][for a review]{Pascucci2023}. Among them we highlight here the [S\,II] $\lambda$4068, [O\,I] $\lambda$6300, and [Ne\,II] 12.81 \micron{} lines. The wind-tracing nature of these lines is inferred by blue-shifts (up to $\sim$ 20 km~s$^{-1}$) in their high-resolution spectra. Among these wind diagnostics, the [O\,I]$\lambda$6300 line is studied in great detail as it is ubiquitously found in disk environments \citep[e.g.,][]{Simon2016,Banzatti2018,Nisini2018,Fang2023b}. For instance, in a survey of forbidden lines in 108 T Tauri stars in NGC 2264, \cite{McGinnis2018} found that 107 of them had an [O\,I] low-velocity gaussian component ($\sim$ 30 km~s$^{-1}$) consistent with a disk wind. In another survey using high-resolution spectroscopy, \cite{Pascucci2020} modeled the [O\,I]$\lambda$6300 low-velocity from several transition disks and showed that most of the emission is coming from within $\sim$1-2\,au. More recently, \cite{Fang2023} spatially resolved the [O\,I]$\lambda$6300 emission from TW~Hya and showed that 80\% comes from within 1\,au of the star, in agreement with previous inferences based on modeling spectrally resolved [O\,I] profiles \citep{Pascucci2011,Pascucci2020}. The small emitting radius in TW~Hya as well as in other sources, combined with blue-shifts in the line centroids, are difficult to reconcile with the PE wind scenario \citep[but see][]{Rab2023} and [O\,I] is therefore more frequently interpreted as tracing an MHD disk wind \citep[e.g.,][]{Simon2016,Banzatti2019,Fang2023b}. 

On the other hand, the [Ne\,II] 12.81 \micron{} detections using \textit{Spitzer} \citep{Espaillat2007,Lahuis2007,Pascucci2007} at the time of discovery were spectrally unresolved. However, follow-up ground-based high spectral resolution observations of [Ne\,II] 12.81 \micron{} demonstrated that it is an important diagnostic of disk winds \citep[e.g.,][]{Herczeg2007,Pascucci2009,Baldovin2012,Sacco2012}. Disks with inner dust cavities show singly peaked [Ne\,II] profiles with modest widths (FWHM$\sim$15-40 km~s$^{-1}$) and small blue-shifts ($\sim$3-6 km~s$^{-1}$) relative to the stellar systemic velocity \citep{Pascucci2009,Sacco2012} which are consistent with the PE wind model predictions \citep{Alexander2008}. Additionally, the absence of the red-shifted component of [Ne\,II] suggests that there is dust in the mid-plane, which blocks our view of the receding wind \citep{Alexander2008,ErcolanoOwen2010,Pascucci2011} and that the emission originates largely outside the dust hole.

However, line profiles alone cannot determine whether [Ne\,II] also traces MHD disk winds. Furthermore, they cannot discriminate between PE winds driven by EUV \citep{Alexander2008} and X-ray radiation \citep{ErcolanoOwen2010}, i.e., wind mass-loss rates that differ by two orders of magnitude. Unlike [O\,I], the [Ne\,II] emission has never been spatially resolved to analyze whether it is tracing emission from the very inner regions of the disk ($<$1-2 au, similar to [O\,I]) or if it is tracing a more extended emission. This is important because PE winds cannot be driven from the very inner regions ($<$1-2au for solar mass stars), which are too deep into the gravitational potential well of the star for the gas to escape \citep[e.g.,][]{Hollenbach1994,Font2004,Clarke2016}. On the contrary,  MHD winds can easily be driven from these very inner regions, close to the star, and can also extend outwards up to tens of au \citep[e.g.,][]{Gressel2015,Bethune2017}. Hence, by observing the radial extent of [Ne\,II] emission in a disk with a very small inner disk and a large dust gap, it can be shown whether the wind is compact or extended. From such a disk, an unresolved, i.e., compact emission, would suggest that [Ne\,II] is tracing an MHD wind. On the other hand, a resolved extended emission would be compatible with both the PE wind scenario as well as radially extended MHD wind models.

To test this hypothesis, we employ MIRI Medium Resolution Spectroscopy (MRS) to observe T\,Cha, a young T Tauri star, surrounded by a disk with a large dust cavity and test whether the known [Ne\,II] emission is spatially extended. The paper is organized as follows: In Section~\ref{sec:obs}, we provide details of the source, observations, and the calibration technique employed to reduce the data. In Section~\ref{sec:results}, we detail the analysis and present our results demonstrating that [Ne\,II] emission is spatially extended in T\,Cha (Sub-section~\ref{sec:extended_wind}). We discuss our results in Section~\ref{sec:discussion} and summarize this work in Section~\ref{sec:sum&con}.

\section{Source, observations, and calibration} \label{sec:obs}

\subsection{Details about the source}

This study focuses on T\,Cha, a bright G8 spectral type star \citep{Chavarria1989,Alcala1993,Schisano2009} located at a distance of $\sim$ 103 pc as calculated using the GAIA Data Release 3 \citep{gaiadr3}, and a member of the $\epsilon$-Cha association \citep[see][for a recent examination of the $\epsilon$-Cha membership]{Murphy2013,Dickson-vandervelde2021}. The estimated age of T\,Cha has a large range and is estimated to be between 2 and 10 Myr \citep{Fernandez2008,Ortega2009}, similar to the age range reported for $\epsilon$-Cha \citep[$\sim 3 -8$\,Myr,][]{Dickson-vandervelde2021}. The primary reason for selecting T\,Cha as a target for our MIRI MRS observation is its strong [Ne\,II] 12.81 \micron{} detection \citep{Lahuis2007} with a small blue-shift in the centroid velocity ($\sim 3 - 6$\,km~s$^{-1}$) and a line width $\sim$ 40 km~s$^{-1}$ indicating that it is tracing a disk wind \citep{Pascucci2009,Sacco2012}. \cite{Pascucci2009} could fit this [Ne\,II] line profile very well with the PE wind model from \cite{Alexander2008} assuming that T\,Cha is a nearly edge-on disk and the gas disk does not have a gap. Later, both assumptions were found to be true: the disk of T\,Cha is viewed at an inclination angle of $\sim$ 70$^\circ$ \citep{Huelamo2015,Pohl2017,Hendler2018} and no gap is detected in the gas disk \citep{Huelamo2015,Wolfer2023}. The absence of a gap is consistent with the fact that the star is accreting, with an accretion rate of $\sim$4 $\times$ 10$^{-9}$ M$_\odot$ yr$^{-1}$ \citep{Cahill2019}. 
In contrast to the gas component, the dust disk has a large gap from $\sim 1-20$\,au and an outer ring from $\sim$ 20-50\,au \citep{Olofsson2013,Pohl2017,Hendler2018}. 

In addition to the high-resolution (R$\sim$600) \textit{Spitzer} IRS spectrum that revealed a strong [Ne\,II] line at 12.81 \micron{} \citep{Lahuis2007}, T\,Cha was also observed in the low resolution (R$\sim$105) mode \citep{Kessler2006}. The Polycyclic Aromatic Hydrocarbon (PAH) feature at 11.2 \micron{} is clearly detected in the low-resolution mode \citep{Geers2006,Kessler2006} while the [Ne\,III] line at 15.55 \micron{} is barely detected in the high-resolution mode \citep{Lahuis2007}. The [Ar\,II] and  [Ar\,III] lines at 6.98 and 8.99 \micron{} are not detected at low resolution, although [Ar\,II] was predicted to be as strong as [Ne\,II] \citep{Hollenbach2009}. 

\subsection{Observations and data calibration details}
T\,Cha was observed as part of the Cycle~1 General Observer program (proposal ID: 2260, PI: I. Pascucci) using the MIRI MRS instrument aboard the James Webb Space Telescope (JWST) \citep{Wells2015}. It was observed on Aug 13-14, 2022, for 3.24 hrs on-source and 4.48 hrs, including the overheads. We used a 4-point dither pattern optimized for an extended source around the central position of T\,Cha - R.A. (J2000) 11:57:13.5269 and DEC. (J2000) -79:21:31.52. The source acquisition was not employed as the aim was to perform spectral mapping of an extended source. The Integral Field Unit (IFU) data was taken in all three bands [SHORT, MEDIUM, LONG] in all 4 channels covering the entire MRS wavelength range of $\sim$ 4.9$-$28\,$\micron$. 

There were 14 integrations per dither point, which sums to a total of 56 integrations for 4 dither pointings. A total of 25 groups were taken in FASTR1 mode per integration, and the integration time per MRS sub-band and complete dither was 4030s. A dedicated background was also observed using identical settings as for T\,Cha. A nearby field was chosen as a background owing to the fields` relative deficiency of astronomical sources using the WISE All-Sky catalog \citep{Wright2010}. 

We used the version `1.11.2' of the JWST calibration pipeline \citep{Bushouse2023}, which was the latest version made available as of July 12, 2023. This version uses the updated MRS spectral response function part of the \texttt{Spec2Pipeline}, and improved cube building and outlier detection step of the \texttt{Spec3Pipeline}. We used the CRDS (Calibration Reference Data System) version `11.16.21' and context `jwst\_1100.pmap' for reference files and their selection rules, respectively. 

The fringe correction step is executed using a fringe flat field and is included in the pipeline as part of the \texttt{Spec2Pipeline}. We further reduced the fringing by applying the \texttt{residual\_fringe} step (Kavanagh et al. in prep.), which is included in the JWST calibration pipeline package as part of \texttt{Spec2Pipeline} but switched off by default. We used the \texttt{outlier\_detection} step part of the \texttt{Spec3Pipeline}, which successfully removes bad pixels from the data. We compared three different methods of background subtraction (a) pixel-by-pixel subtraction before cube creation as part of the \texttt{Spec2Pipeline}, (b) uniform model background subtraction as part of the \texttt{Spec3Pipeline}, and (c) pixel-by-pixel subtraction after cube creation: post-pipeline. Among these, method (c) reduced fringing at longer wavelengths as well as the number of outliers, and hence we adopted method (c) for our background subtraction. Looking at the IFU (Integral Field Unit) cubes, we find that T\,Cha is not centered in the FOV (Field Of View). This is because the effect of the proper motion from J2000 coordinates was not accounted for during the observation. However, we safely retain all the source emission in all four channels owing to which, we obtain a spectrum for T\,Cha for the complete wavelength range of MIRI MRS ($\sim$5-28 \micron{}). After creating a pixel-by-pixel background subtracted IFU cube, we retrieve a 1-D spectrum by summing all the pixels in an aperture of radius 2.5 $\times$ FWHM, where FWHM = $0.033 \times(\lambda/\micron{}) + 0".106$ \citep{Law2023}. We perform the aperture correction after extracting the spectrum, with values retrieved from the aperture correction reference file in CRDS. 

One of the key science goals of our program was to evaluate if emission in lines known to trace disk winds is extended. As such, in addition to inspecting 1D spectra, we perform spaxel-by-spaxel continuum subtraction (Section~\ref{sec:contsub}) as well as PSF (Point Spread Function) subtraction (Section~\ref{sec:psfsub}) in selected emission lines. We use the standard star HD37962, which is a bright G2V star, as a proxy for the PSF. HD37962 was observed under the proposal id 1538, also noted as CAL-FLUX-003, which is an absolute flux calibration program \citep{Gordon2022}. HD37962 was observed with a 4-point dither, similar to T\,Cha, for $\sim$ 2.2 hrs including the overheads. A dedicated background was also observed for HD37962 for $\sim$ 0.7 hrs using a 2-point dither. Both observations covered the whole wavelength range of MIRI MRS ($\sim$ 5-28 \micron{}) and used the FAST readout mode. We obtained the uncalibrated raw files for both observations using the Barbara A. Mikulski Archive for Space Telescopes (MAST\footnote{The Mikulski Archive for Space Telescopes (MAST) is a NASA-funded project to support and provide a variety of astronomical data archives, with the primary focus on the optical, ultraviolet, and near-infrared parts of the spectrum. MAST is located at the Space Telescope Science Institute (STScI).}) and calibrated them using the same methodology as for T\,Cha, except for the background subtraction step. As HD37962 and its background are observed in different dither configurations, we could not perform pixel-by-pixel subtraction and instead performed the master background subtraction which is available as part of the \texttt{calspec3\_pipeline}.

\begin{figure*}
    \centering
    \includegraphics[width=\textwidth]{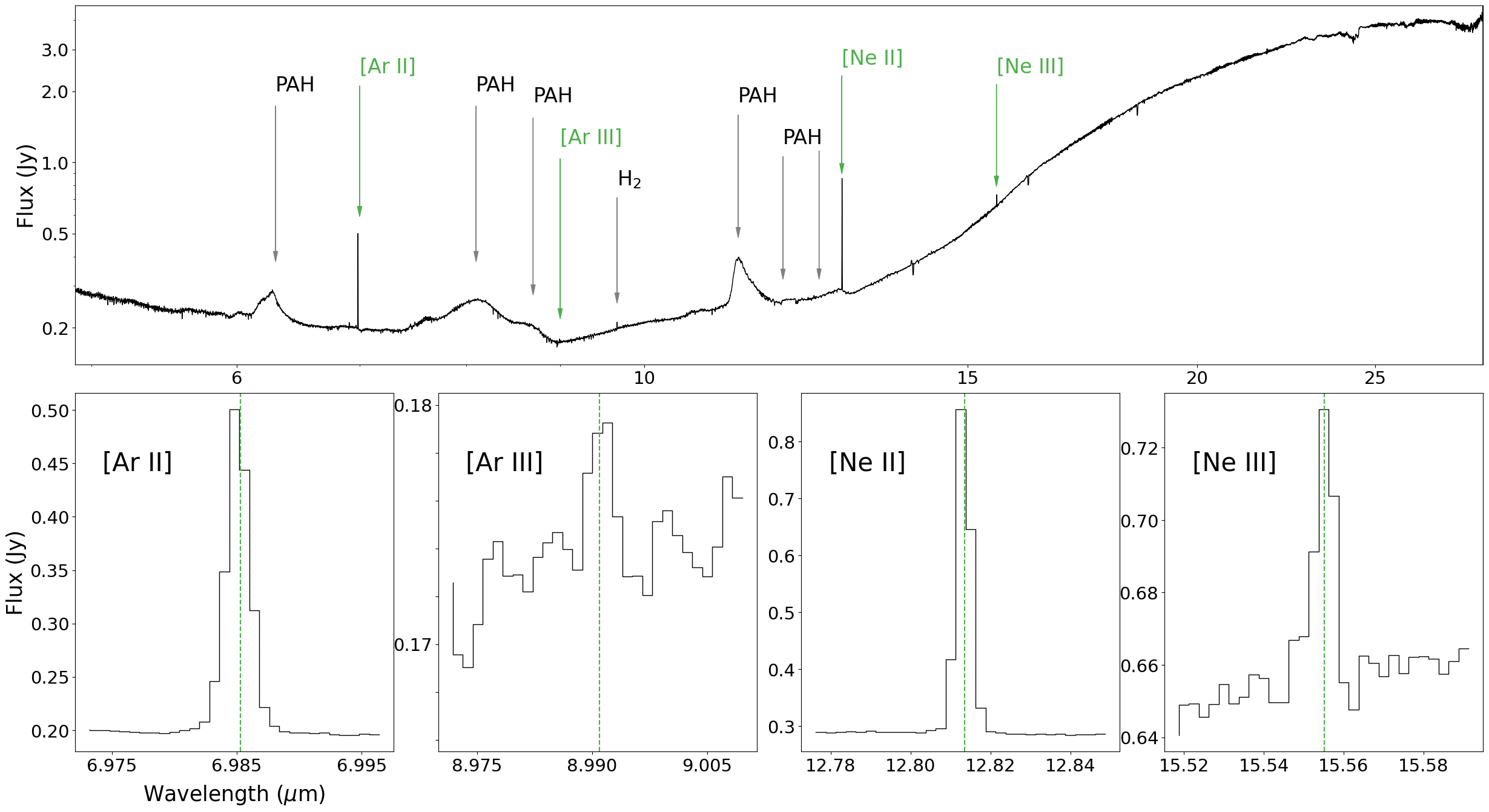}
    \caption{The upper panel is the JWST MIRI MRS spectrum of T\,Cha plotted between $\sim$ 5 \micron{} and $\sim$ 28 \micron{} with the PAH features and H$_2$ $\nu$0-0 S(3) line marked in black and the forbidden noble gas emissions in green. The lower four panels further highlight the four forbidden line emissions: [Ar\,II], [Ar\,III], [Ne\,II], and [Ne\,III] which will be the focus of this study.}
    \label{fig:spectra}
\end{figure*}

\begin{table*}
    \centering
    \caption{Measured line fluxes, 3$\sigma$ upper limits and continuum level}
    \begin{tabular}{c c c c c | c c}
    \hline
    Line & Wavelength$^a$ & \multicolumn{2}{c}{This work} & Published & \multicolumn{2}{c}{Continuum level$^c$}\\ 
    &  & \multicolumn{1}{c}{JWST} & \multicolumn{1}{c}{Spitzer$^b$} & Spitzer & \multicolumn{1}{c}{JWST} & \multicolumn{1}{c}{Spitzer}\\ 
    & (\micron{}) & \multicolumn{3}{c}{(10$^{-14}$ erg s$^{-1}$ cm$^{-2}$)} & \multicolumn{2}{c}{(Jy)} \\ \hline
    [Ar\,II] & 6.98 & 4.31 $\pm$ 0.02 & $<$10.5 & $<$10.68$^d$ & 0.2 & 0.69\\ \newline
    [Ar\,III] & 8.99 & 0.08 $\pm$ 0.02 & $<$0.81 & \nodata & 0.175 & 0.455\\ \newline
    H$_2$\,S(3) & 9.66 & 0.13 $\pm$ 0.01 & $<$0.8 & \nodata & 0.195 & 0.41\\ \newline
    HI\,(7-6) & 12.37 & $<$0.15 & $<$0.9 & \nodata & 0.27 & 0.32\\ \newline
    [Ne\,II] & 12.81 & 4.99 $\pm$ 0.01 & 3.2 $\pm$ 0.2 & 3.2 $\pm$ 0.2$^e$ & 0.29 & 0.3\\ \newline
    [Ne\,III] & 15.55 & 0.52 $\pm$ 0.02 & $<$0.5 & $<$0.2$^e$ & 0.655 & 0.3\\ \hline
    PAH$^{f}$ & 6.2 & 95.4 & \nodata & \nodata & 0.23 & 0.82\\ 
    PAH$^{f}$ & 11.2 & 104 & 41 & 33$^{g}$ & 0.26 & 0.38\\ 
    \hline
    \end{tabular}
    \tablecomments{The reported 1$\sigma$ uncertainty for the JWST spectra should be added in quadrature with the absolute flux error \citep[$\sim$5\%,][]{Argyriou2023}}
    \tablenotetext{a}{Rest wavelength of the line}
    \tablenotetext{b}{Recalculated values in this work using the \textit{Spitzer} IRS high-res spectrum (R$\sim$700)}
    \tablenotetext{c}{Continuum flux at the rest wavelength of the corresponding lines}
    \tablenotetext{d}{Published value from \cite{Szulagyi2012}}
    \tablenotetext{e}{Published values from \cite{Lahuis2007}}
    \tablenotetext{f}{The flux is calculated as the area under the feature without fitting a Gaussian profile}
    \tablenotetext{g}{Published value from \cite{Geers2006}}
    \label{tab:intensities}
\end{table*}

To take into account the MIRI MRS PSF variations and their impact on our results, we perform the PSF comparison and subtraction with an additional standard star HD167060 which is a G3V star and was observed the same day as T\,Cha: August 14, 2022. HD167060 was observed under the same program (1538) as HD37962 for $\sim$ 2.8 hrs with a dedicated background observed for $\sim$ 0.6 hrs including the overheads. The observation settings and data reduction technique employed for HD167060 are the same as for HD37962.

\section{Analysis and Main Results} \label{sec:results}

Here we discuss the detection of four forbidden noble gas lines with the first-ever detection of [Ar\,III] in a protoplanetary disk. We also describe the two independent methods employed for discovering the spatially extended emission in [Ne\,II].

\subsection{Detection of multiple transitions from ionized noble gases} \label{sec: lines}

In accordance with previous observations, we have a strong detection of the [Ne\,II]  12.81 \micron{} line in the JWST MIRI-MRS spectrum of T\,Cha. We also detect the [Ne\,III] 15.55 \micron{}, [Ar\,II] 6.98 \micron{} and [Ar\,III] 8.99 \micron{} emission lines as shown in Figure~\ref{fig:spectra} making it the first-ever spectrum of a protoplanetary disk to have detected four forbidden noble gas lines together. Additionally, we also detect PAH bands at 6.02 \micron{}, 8.22 \micron{} as well as PAH complexes at 6.2 \micron{}, 7.7 \micron{}, 8.6 \micron{}, 11.3 \micron{}, 12 \micron{}, and 12.7 \micron{}, of which only the 11.2 \micron{} complex was detected for T\,Cha using \textit{Spitzer} \citep{Geers2006,Kessler2006}. Furthermore, we detect the H$_2$$\nu$0-0 S(3) line at 9.66 \micron{} and do not detect the most commonly detected hydrogen recombination line HI (7-6) at 12.37 \micron{} \citep[e.g.,][]{Rigliaco2015}.
 
To calculate line fluxes and their uncertainties we adopt a Monte-Carlo approach. First, we fit a straight line to $\sim$20 data points outside the transition on either side and calculate the standard deviation relative to the fitted line. Then, a set of 10,000 spectra is generated within the range of a Gaussian distribution with a sigma for each point equal to the previously calculated standard deviation. For all of the spectra, we fit a Gaussian profile and compute the area under the curve. We use the mean and standard deviation of these areas as the final flux and the associated uncertainty, respectively \citep[e.g.,][]{Pascucci2008, Banzatti2012}. To calculate the upper limits in case of non-detections, we use the equation:

\begin{equation}
    F_{up} = 3\sigma d_{\lambda}\sqrt{N}
\end{equation}

Where $\sigma$ is the root mean square (rms) of the data points where the line would have been, and the line flux is computed over the spectral resolution element (N = 2 pixels), assuming the noise is uncorrelated. Here, $d_{\lambda}$ is the wavelength of the given line divided by the resolution \citep[see][for a similar approach]{Szulagyi2012}. These integrated fluxes and upper limits from the MIRI spectrum of T\,Cha are listed in Table~\ref{tab:intensities} along with the estimates from \textit{Spitzer} IRS spectrum, calculated using the same technique. We do not detect HI (7-6), and provide an upper limit in Table~\ref{tab:intensities}. The non-detection of the HI (7-6) line aligns with the low HI (7-6)/[Ne\,II] ratio of 0.008 expected for disk winds \citep[see][]{Hollenbach2009}. Even in the scenario where HI (7-6) traces accretion \citep[see][]{Rigliaco2015}, T\,Cha has a weaker emission than average given its accretion luminosity. This may be due to the HI emission being partly obscured by the nearly edge-on disk of T\,Cha, which was also suggested to explain low H$\alpha$ emission by \cite{Cahill2019}. A Gaussian fit to each of the lines listed, including HI (7-6), is described in Section \ref{sec:gauss_fit} in the Appendix.

Using the method described above, we report a 4$\sigma$ detection of [Ar\,III], accompanied by a $>$10$\sigma$ detection of both [Ne\,III] and H$_2$ S(3). The most robust detections, [Ar\,II] and [Ne\,II] exhibit signals exceeding 200$\sigma$. Owing to the strong detections of [Ar\,II] and [Ne\,II], we focus on exploring any spatial extension in these lines.

\subsection{Spatially extended [Ne\,II] wind emission}
\label{sec:extended_wind}

As mentioned in Section \ref{sec:obs}, high-resolution ground-based spectroscopy demonstrated that the [Ne\,II] at 12.81\,\micron{} from T\,Cha is slightly blue-shifted ($\sim$4 km~s$^{-1}$) relative to the stellar velocity, hence likely tracing a slow disk wind \citep{Pascucci2009, Sacco2012}. Here, we further show that the [Ne\,II] emission is spatially extended by employing two different techniques to isolate the line emission and compare it with the PSF.

\subsubsection{Spaxel-by-Spaxel Continuum Subtraction}
\label{sec:contsub}

In this first approach, we remove the continuum in each spaxel to generate line-only maps. To do this, we consider $\sim$ 30 continuum data points on both sides of the line emission and curve-fit an n-degree polynomial, such that n is the lowest integer that best fits the data. In the case of T\,Cha, we can use a 2-degree polynomial. For wavelength points in a spaxel under consideration, we subtract a 2-order polynomial continuum model flux and create a corresponding continuum-subtracted cube. Simultaneously, using the polynomial continuum model flux for each wavelength point in a spaxel, we create a continuum emission cube too \citep[see][for a similar method applied to H$_2$ rovibrational lines]{Beck2019}. We repeat this procedure and re-generate the continuum and continuum-subtracted datacubes for each line individually. Each time, our continuum fit is good only for the $\sim$ 30 points considered on both sides of the line and likely poor anywhere beyond it. We apply this procedure to all three strongly detected forbidden lines (except [Ar\,III] owing to its relatively weak detection of $\sim$ 4$\sigma$) to create their continuum-subtracted maps. 

\begin{figure*}
    \centering
    \includegraphics[width=1.5\columnwidth]{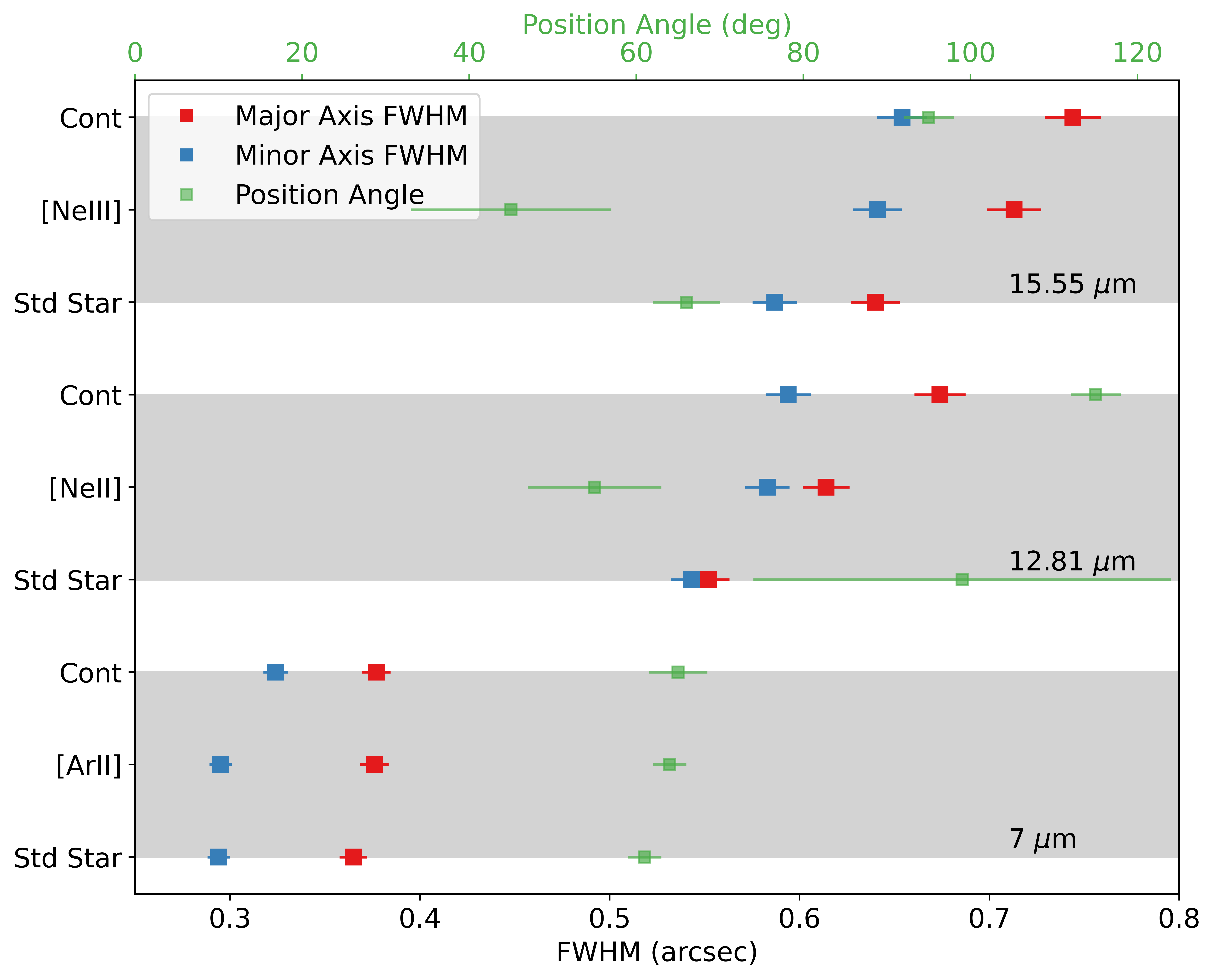}
    \caption{Comparison of the 2-D Gaussian size calculated from images of the continuum-subtracted line, T\,Cha continuum (represented as \emph{Cont}) and the standard star (represented as \emph{Std Star}). The three are compared at the same wavelength for [Ar\,II], [Ne\,II], and [Ne\,III] respectively.}
    \label{fig:size_comparison}
\end{figure*}

For each of these continuum-subtracted line maps, we select the wavelength plane with the peak line emission and fit a 2-D Gaussian to determine the size and Position Angle (PA) of the emission. We also fit a 2-D Gaussian to the corresponding wavelength plane in each of the continuum maps. We perform this 2-D Gaussian fitting using the \texttt{imfit} task which is part of the Common Astronomy Software Application \citep[CASA,][]{McMullin2007} package. We apply the same approach to the standard star (HD37962) at the peak line wavelengths of [Ar\,II], [Ne\,II], and [Ne\,III] lines in T\,Cha to compare the component size for possible extension beyond a point source. HD37962 was favored over HD167060 owing to its comparatively higher luminosity. Results from our analysis are shown in Figure~\ref{fig:size_comparison}. Of the three wavelengths shown, the size of the standard star emission (PSF) is smallest at 7\,\micron{} and largest at 15.55\,\micron{}, as expected since the PSF size increases linearly as a function of wavelength, see the relation in Section \ref{sec:obs}. At 7\,\micron{}, the size and PA of the standard star, and the continuum-subtracted [Ar\,II] are similar within the error bars suggesting compact [Ar\,II] emission. Similarly, the major axis of the nearby continuum and the PA are the same within the error bars suggesting a compact emission too, however, the minor axis is slightly larger than the PSF hinting at a slightly resolved 7\,\micron{} emission.

At 12.81\,\micron{}, the standard star is more circular and therefore has a large error bar associated with its PA. It is clear that the [Ne\,II] emission is more extended/resolved than the standard star and the continuum is further extended than the [Ne\,II] emission. Note that the continuum PA at 12.81\,\micron{} is 115$\pm$ 3$^\circ$ which is in strong agreement with the previous estimates of the disk PA: 113 $\pm$ 6$^\circ$ \citep[][using ALMA CO (3-2) integrated emission map]{Huelamo2015}, $\sim$113$^\circ$ \citep[][using ALMA dust continuum at 3mm]{Hendler2018}, and $\sim$ 114$^\circ$ \citep[][using VLT/SPHERE observations of the dust disk]{Pohl2017}. Importantly, not only is the [Ne\,II] emission more extended than the standard star, but it also fits a significantly different PA than the T\,Cha continuum, signaling that [Ne\,II] is tracing emission away from the disk. This further corroborates the disk wind hypothesis inferred from  the spectrally resolved [Ne\,II] profiles. At 15.55\,\micron{}, we see similar results as for [Ne\,II]. Specifically, the [Ne\,III] emission is significantly extended beyond the standard star, and its PA is the same within the error bars as the [Ne\,II] and different from the continuum. The continuum too at 15.55\,\micron{} is extended beyond the standard star as would be expected considering the extension in the 12.81\,\micron{} continuum. We note that the disk PA inferred from the 15.55\,\micron{} continuum emission does not agree with that at 12.81\,\micron. We speculate that this may be due to the degradation in spatial resolution at longer wavelengths, such that the extension in the continuum emission is less reliable at 15.55 than at 12.81\,\micron.

It is also interesting to note that when this same method is applied to the PAH features, particularly the one centered around 6.2 \micron{}, we find these features, too, are more extended compared to the standard star. We find that these PAH features are more extended than the continuum at that wavelength, too - the Gaussian size of the 6.2 \micron{} PAH emission is 0.4" $\times$ 0.36" while that of the standard star and continuum is 0.35" $\times$ 0.28".

\subsubsection{PSF subtraction}
\label{sec:psfsub}

\begin{figure*}
    \centering
    \includegraphics[width=\textwidth]{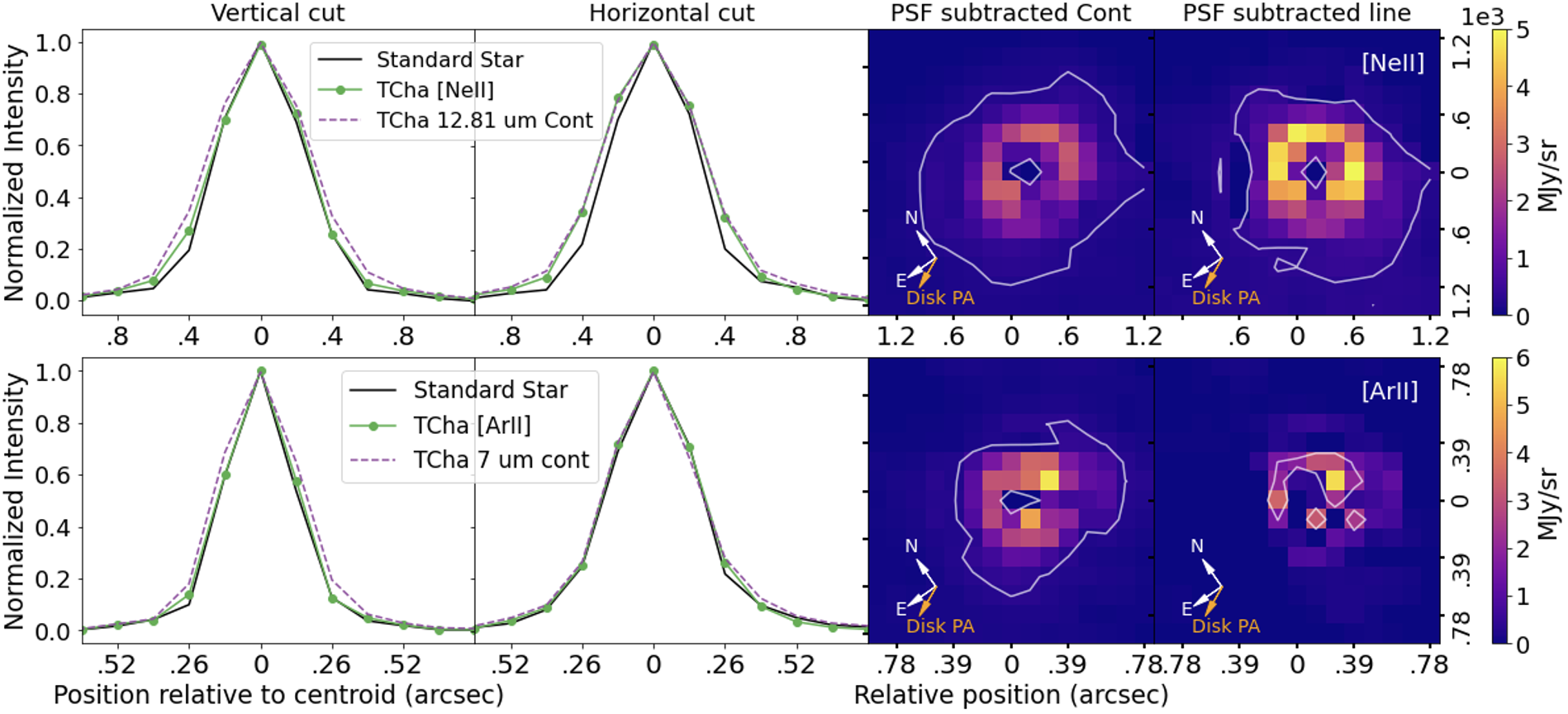}
    \caption{The left two panels compare the radial profiles of continuum and continuum-subtracted lines with the PSF (standard star HD37962) in the horizontal and vertical directions. The two panels on the right show the PSF-subtracted images of the continuum at 12.81 and 7 \micron{} and [Ne\,II] and [Ar\,II] lines. The white contours in the PSF-subtracted images follow delineate 5$\times$ the rms calculated in the source-emission free regions in the PSF subtracted images.}
    \label{fig:psfsub}
\end{figure*}

We perform PSF comparison as well as subtraction to test the results reported in the previous sub-section independently. When we conduct PSF subtraction on the [Ne\,II] line emission, we find it to be spatially resolved in agreement with the results from Gaussian fitting. We treat the same standard star observation used in the previous section, HD37962, as the PSF. To effectively compare our observations with the PSF, we reduce the observations in the detector plane using the `ifualign' mode during cube building. Though T\,Cha is off-centered in the detector plane, it does not affect our routine, as no field-dependent distortion has been seen for the PSF of MIRI MRS (Patapis, P., personal communication, 2023) so far. To perform PSF comparison, we first find the image centroid for the line image as well as the PSF using the \texttt{centroid\_2dg} function part of Photutils.\footnote{Photutils is an Astropy package for detection and photometry of astronomical sources \citep{larry2022}}. Using the centroid, both sources are shifted such that the centroid lies at the center of their respective images. This step is carried out using the \texttt{interpolation.shift} utility of the \texttt{scipy.ndimage} package, which is very well-tested in a broad range of applications. The interpolation function used is bicubic spline. Next, we scale the PSF emission to the line emission by using a weighted average ratio of a 5 $\times$ 5-pixel grid around the central spaxel.

We compare the scaled PSF with the continuum and line emission at 12.81 \micron{} and 7 \micron{} by drawing radial profiles in horizontal and vertical directions passing through the central bright pixel (see the left half of Figure~\ref{fig:psfsub}). We try to perform the same analysis on [Ne\,III] as well, however, the emission in the continuum subtracted image is not strong enough to find its centroid with confidence. Hence, we restrict our analysis to [Ne\,II] and [Ar\,II]. At 12.81 \micron{}, both the vertical and horizontal profiles of the continuum as well as [Ne\,II] show extension beyond the PSF profile. However, at 7 \micron{}, only the continuum is slightly extended beyond the PSF in the vertical profile. The [Ar\,II] emission in both the vertical and horizontal profiles closely follows the PSF. It should be noted that, in both cases, the continuum is more extended than the line emission in the vertical direction as expected since the disk orientation is close to being vertical in the image plane considering its PA of $\sim$ 115 deg from the North (N) shown in Figure \ref{fig:psfsub}.

To corroborate these results using the 2D image, we subtract the scaled PSF image from the continuum and line images, respectively. These PSF-subtracted images at 12.81 \micron{} and 7 \micron{} are shown in Figure~\ref{fig:psfsub} (right half). The white contours trace 5$\times$ the rms calculated in a PSF-sized aperture away from the source in the PSF-subtracted images. The contours clearly show that [Ne\,II] traces extended emission beyond the PSF. As can be seen, the continuum at 12.81 \micron{} is also well resolved and shows a broad emission. It is important to note that, when looked at closely, the direction of extension in [Ne\,II] is not the same as the direction of extension in the continuum image suggesting that [Ne\,II] is tracing an emission away from the disk, which is compatible with a wind. This is much more clearly seen in the difference of their PA in Figure~\ref{fig:size_comparison}, as also mentioned earlier. To estimate the component size of [Ne\,II] emission, we subtract the PSF FWHM in quadrature with [Ne\,II] FWHM in both horizontal (wind) and vertical ($\sim$disk) directions. This results in 0.32" in the wind direction and 0.13" in the disk direction which translates to $\sim$33 au and $\sim$13 au at a distance of 103 pc. At 7 \micron{} however, the continuum is only slightly resolved, and [Ar\,II] very clearly traces a much more compact emission as was also seen through the vertical and horizontal curves investigated earlier and through 2-D Gaussian fit to continuum subtracted image in Figure~\ref{fig:size_comparison}.

We also demonstrate that our results are independent of the PSF variations by following the exact same procedure with the standard star HD167060. A similar analysis to Figure~\ref{fig:psfsub} is shown in Figure~\ref{fig:psf2sub} in the Appendix, but for HD167060. It is clear that at 12.81 \micron{} both the continuum and [Ne\,II] are extended in different directions, and at 7 \micron{}, [Ar\,II] shows compact emission, consistent with the results described above. See Appendix \ref{sec:psf2sub} for details on PSF subtraction with HD167060 as the standard star.

Additionally, we perform this method on the 6.2 \micron{} PAH feature and find it to be spatially extended, in agreement with the result from the continuum subtraction presented in the previous sub-section. We present this result for 6.2 \micron{} PAH in Appendix \ref{sec:extended_pah}.

\section{Discussion} \label{sec:discussion}

Through the MIRI MRS observation of T\,Cha, we found several forbidden noble gas lines, which possibly trace disk winds, and many PAH emission bands along with the H$_2$$\nu$0-0 S(3) line. Among these lines, [Ne\,II] has been known as a disk wind tracer for over a decade \citep[e.g.,][]{Pascucci2009}. We found that the JWST flux of [Ne\,II] in T\,Cha is higher than that of Spitzer. Additionally, the infrared continuum of T\,Cha has also changed considerably as evident from Table \ref{tab:intensities}, which prompts us to contextualize the two results in Section \ref{sec:line_var}. We also use line ratios to determine the main ionization source of these lines in Section \ref{sec:line_ratio}.

Our analysis in Section \ref{sec:extended_wind} revealed that the [Ne\,II] and [Ne\,III] emission is extended beyond the PSF while [Ar\,II] is not. In Section \ref{sec:spatial_extent} we discuss these results in the context of PE and MHD winds. Finally, in Section \ref{sec:PAH} we discuss the nature of the extended PAH emission we reported in Section \ref{sec:extended_wind} 

\subsection{The origin of continuum and line variability} 
\label{sec:line_var}

The JWST and \textit{Spitzer} T\,Cha continuum values in Table \ref{tab:intensities} clearly demonstrate that the disk is variable at infrared wavelengths. While the continuum flux density at $\sim 7$\,\micron\ has decreased by a factor of $\sim 3$ from 2006 to 2022, that at $\sim 16$\,\micron\ has increased by a factor of $\sim 2$ (see Xie et al. in prep, for more details on the continuum variability). This type of variability has been reported for other disks \citep[e.g.,][]{Espaillat2011} but it is typically smaller than that observed for T\,Cha. This means that for T\,Cha, during the JWST observations, there is less dust attenuation and more stellar radiation could penetrate the inner disk and reach the outer disk than during the \textit{Spitzer} time.  

Interestingly, the pivot wavelength\footnote{The wavelength at which there is no variability} of the seesaw variability is around 12.8\,\micron{}. While the continuum near the [Ne\,II] line has not changed, the line flux has increased from the Spitzer time by $\sim$50\% (see Table~\ref{tab:intensities}). This cannot be attributed to differences in the apertures used for spectral extraction because they are both large enough to encompass the entire emission -- 3.6" for IRS and 2.5" diameter circle for MIRI-MRS. The change in the [Ne\,II] flux is not as large as that observed by \cite{Espaillat2023} in SZ\,Cha ($\sim$150\%) which, combined with the [Ne\,III] variability, was reasoned as a switch from EUV ionization to X-ray ionization of Neon. However, this does not seem to be the case for T\,Cha as its [Ne\,II]/[Ne\,III] ratio was $>$1 even during the Spitzer time (Table \ref{tab:intensities}), indicating ionization by X-rays (Section \ref{sec:line_ratio}). The increase in T\,Cha's [Ne\,II] emission might result from several factors, e.g. increase in the wind mass loss rate, more heating, or higher fractional ionization. The latter two options are in line with the reduction of the inner disk mass (Xie et al., in prep) and the [Ne\,II] emission tracing emission out to the outer disk. However, a caveat with more heating is that it will enhance the emission only if the wind is colder than the excitation temperature ($<$1200 K).

T\,Cha is also highly variable at optical wavelengths \citep[up to 3\,mag in V-band,][]{Covino1992,Alcala1993,Walter2018}, and so is the [O\,I] $\lambda$6300 line \citep[e.g.,][]{Covino1996}. Dedicated monitoring has identified two different timescales for the [O\,I] variability: a short-term variability \citep[$\sim$1-2 days,][]{Schisano2009}; and a relatively long-term variability \citep[$\leq$ 2 years,][]{Cahill2019}. \cite{Schisano2009} detected the short-term variability using three independent observation runs between 1993 and 1995 with observations in each run separated by less than a day. They found the variability timescale comparable to the optical photometric variations and also a positive correlation with the visual extinction. This led them to propose an ``occulting clump" with a period of a few days and a maximum visual extinction magnitude $\sim$ 3.6 as a possible explanation for the optical continuum and line variabilities. On the other hand, \cite{Cahill2019} reported that the average [O\,I] 6300 line flux dropped from 1 $\pm$ 0.1 $\times$ 10$^{-13}$ erg cm$^{-1}$ s$^{-2}$ recorded in 2012 to 0.3 $\pm$ 0.4 $\times$ 10$^{-13}$ erg cm$^{-1}$ s$^{-2}$ recorded in 2014 suggesting a relatively ``long-term" variability. This ``long-term" variability cannot be explained by the occulting clump hypothesized by \cite{Schisano2009}.

It is possible that the [Ne\,II] variability is similar to [O\,I] ``short-term" and/or [O\,I] ``long-term" variability, however, a visual extinction of $\sim$3.6 mag will be too small at $\sim$13 \micron{} to produce the [Ne\,II] variability making it more likely to be similar to the [O\,I] ``long-term" variability. But having only two data points i.e. \textit{Spitzer} in 2006 and JWST in 2022, makes it difficult to derive any conclusion at present. Moreover, the non-alignment between the observation years of [O\,I] ($\sim$1993, 2012 and 2014) and [Ne\,II] (2006 and 2022) further complicates comparing the evolution of these two lines relative to the disk evolution models of T\,Cha. Simultaneous monitoring of the two lines will be necessary to understand whether the [Ne\,II] and [O\,I] variabilities are correlated or not, the time period of each, and hence, the underlying mechanism(s) for the variabilities. 

\subsection{Constraints on the high-energy photons impinging at the disk surface}
\label{sec:line_ratio}

\begin{figure*}
\centering
\begin{minipage}{.49\textwidth}
  \centering
  \includegraphics[width=\columnwidth]{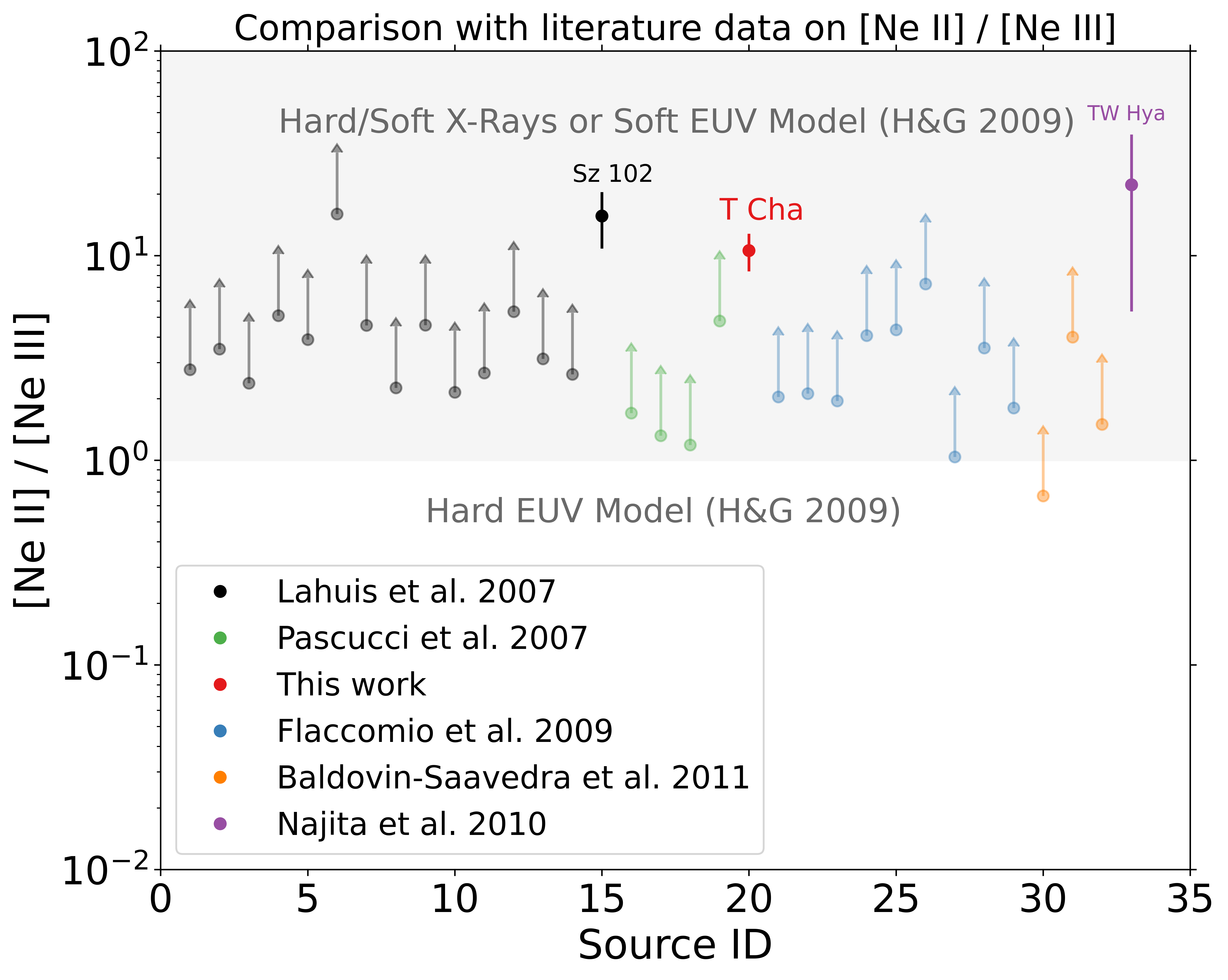}
  \label{fig:test1}
\end{minipage}%
\begin{minipage}{.49\textwidth}
  \centering
  \includegraphics[width=\columnwidth]{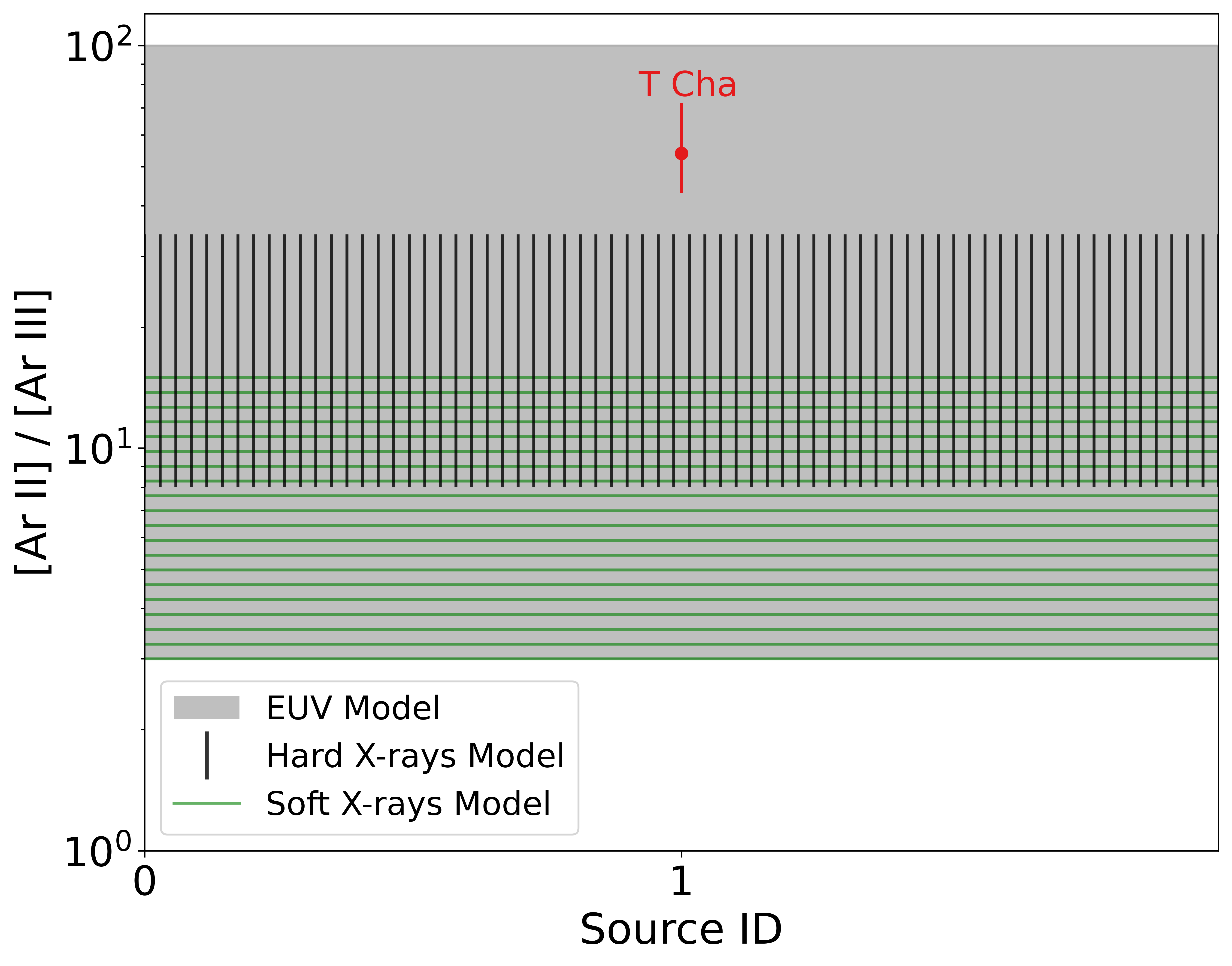}
\end{minipage}
\caption{The left panel shows the [Ne\,II]/[Ne\,III] ratio where the reference for each point is given in the legend, and the right panel shows the [Ar\,II]/[Ar\,III] ratio for T\,Cha. Upward arrows indicate the lower limits, and H\&G 2009 refers to \cite{Hollenbach2009}. The white region in the left panel refers to the Hard EUV Model, which can easily be ruled out as the ionization source for Neon. However, for Argon, the right panel suggests that the EUV model is preferred over X-rays for its ionization. Here, the EUV model combines both Soft and Hard EUV radiation.}
\label{fig:lineratio}
\end{figure*}

The Ne$^+$ and Ar$^+$ ions detected in disks can only be ionized by stellar EUV (13.6 eV $< h \nu <$ 100 eV) and/or X-ray (0.1 keV $< h \nu <$ 10 keV) photons owing to their high ionization potentials (21.56\,eV for Neon and 15.76\,eV for Argon). It is crucial to determine the main source (EUV or X-ray) of disk heating and ionization as it will constrain the wind mass loss rate which in turn will set a limit on the gas-disk lifetime. [Ne\,II]/[Ne\,III] and [Ar\,II]/[Ar\,III] line intensity ratios can help discriminate between cases in which Soft/Hard EUV or Soft/Hard X-ray stellar photons provide the dominant source of disk atmosphere ionization \citep{Hollenbach2009,Szulagyi2012,Espaillat2013}. A “Hard” spectrum is one in which there are more photons at higher energies beyond a certain energy value. Following the literature \citep[e.g.,][]{Hollenbach2009}, we define a Hard X-ray spectrum as that in which there is substantial contribution from photons at and beyond 1 keV, while a Soft X-ray spectrum as that in which flux dominates at 200-300 eV and declines at 1 keV. For EUV, we consider a Hard EUV spectrum as that in which $L(\nu) \propto \nu^{-1}$, whereas Soft EUV as a black-body with $T \sim 20,000-30,000$\,K.

We plot the [Ne\,II]/[Ne\,III] and [Ar\,II]/[Ar\,III] line ratios for T\,Cha in the context of previous \textit{Spitzer} line detections and upper limits (see Figure~\ref{fig:lineratio}). We overplot model predictions from \cite{Hollenbach2009} for [Ne\,II]/[Ne\,III]. For predicting the [Ar\,II]/[Ar\,III] ratios, we used a TW~Hya template stellar spectrum \citep{Ruaud2019} and models of \cite{Hollenbach2009}. T\,Cha is the only object for which there are both ratios. While the [Ne\,II]/[Ne\,III] plot only has three data points with both lines detected, the lower limits are sensitive enough to eliminate the possibility of Hard EUV photons as the dominant Neon ionization mechanism for all but one source.
This is because the line luminosity ratio reflects the ion abundance ratios \citep[see][]{Hollenbach2009}, given the nearly similar excitation energies and A-values. Therefore, [Ne\,II]/[Ne\,III] $\sim$ f(Ne$+$)/f(Ne$^{++}$) where f denotes the fraction of Neon atoms in the respective ionization state. It is easier to produce [Ne\,III] over [Ne\,II] with hard EUV spectra, driving the ratio below one, contrary to the observations. Soft EUV spectra (with T$<$30,000K) have few 41eV photons to ionize Ne+ and therefore result in lower [Ne\,III] emission. X-ray Auger ionization leads to multiply-charged ions, which recombine rapidly and exchange charge with H atoms, leading to high Ne+/Ne++ ratios \citep[see][]{Glassgold2007} for both hard and soft spectra as long as $\sim$1keV photons are present.

T\,Cha (this work) is the only source with both [Ar\,II] and [Ar\,III] detected. The line luminosity ratio is again approximately given by the ratio of the ion abundances to within a factor $\lesssim 2$, and the observed line ratio therefore suggests singly ionized Argon is nearly 50 times more abundant than doubly ionized Argon. However, the ionization energies of Ar and Ar$^+$ are 15.76\,eV and 27.6\,eV respectively and hence not very different. The high line ratio can only be reconciled for high electron densities (lowering the recombination rate of Ar$^{++}$) and a very soft EUV field with $T \sim 20,000$\,K. While model-calculated line ratios favor EUV production versus X-ray production (see Fig.~\ref{fig:lineratio}) of Ar ions, detailed models are needed for a more definitive conclusion.

The cm emission in excess of the dust thermal emission from the disk of TCha provides an upper limit on the EUV luminosity impinging on the disk. This upper limit for T\,Cha is not stringent enough to exclude ionization of Neon from Soft EUV photons \citep{Pascucci2014}. This leads us to ask whether Soft EUV can ionize both Argon and Neon in the disk of T\,Cha? Since EUV heated gas is at $\sim$10,000K, the gas temperature should be high even in the outer disk. As [Ne\,II] and [Ar\,II] have similar I.P's, they should have had similar spatial extent, which is counter to what we find for T\,Cha, i.e. [Ne\,II] and [Ne\,III] are spatially more extended than [Ar\,II] (Section \ref{sec:extended_wind}). Furthermore, it seems unlikely for Soft EUV photons to be able to penetrate (go unabsorbed) through the compact wind component being traced by Argon and ionize Neon further out in the disk. It is particularly difficult for Soft EUV to produce [Ne\,III] which has a high ionization potential ($\sim$41 eV), and is found to be spatially more extended than Argon in T\,Cha. Therefore, we conclude that the most likely scenario is that soft EUV ionizes Argon, while Neon is ionized by X-rays (either hard or soft spectrum), see sketch in Figure \ref{fig:illustration}. This is in line with the higher penetration depths of X-rays with respect to EUV \citep{Hollenbach2009} and the scenario put forward in \cite{Pascucci2020} for more evolved disks where any inner wind becomes more tenuous. Here, we assumed that like [Ne\,II], [Ar\,II] is also tracing a disk wind component, as the [Ar\,II] 6.98 $\micron{}$ line is inaccessible with a high-resolution spectrograph to search for the signature line blue-shift of a wind.

Sellek et al. (submitted) modeled winds in T\,Cha with EUV and X-ray spectra and attempted to constrain the spectrum using the line ratios and line luminosities obtained from this work. They find that adding a Soft EUV component to the X-ray spectrum leads to a better match with the observed ratios of Neon and Argon, which is in agreement with the results from this work.

\subsection{The spatial extent of disk winds traced by Neon and Argon}
\label{sec:spatial_extent}

As mentioned earlier, it is crucial to determine the origin of [Ne\,II] emission (MHD vs. PE wind) and, consequently, the wind mass loss rate to constrain the time left for planet formation and migration. The spatial information on [Ne\,II] emission obtained in this work (Section \ref{sec:extended_wind}) can provide us with useful constraints on its origin as well as the wind mass loss rate estimates. To re-iterate, T\,Cha has a large dust gap \citep{Hendler2018} and its [Ne\,II] emission shows a blue-shift \citep{Pascucci2009,Sacco2012}. This means that our view of the red-shifted emission is blocked either by the very small inner dust disk ($<$1 au) or the outer disk ($\sim$25 au) \citep[e.g.,][]{ErcolanoOwen2010,Pascucci2011}. If the [Ne\,II] emission was spatially compact, then it would have eliminated its possibility of tracing a PE wind, but as we have seen in both Sections~\ref{sec:contsub} and \ref{sec:psfsub}, the [Ne\,II] emission is extended beyond the PSF and hence, resolved. Moreover, we also see that [Ne\,III] is similarly extended beyond the PSF from Section \ref{sec:contsub}. The difference in the extension direction between the continuum emission at 12.81 \micron{} and [Ne\,II] emission provides spatial confirmation for [Ne\,II] tracing a disk wind. More importantly, this extension is compatible with the PE wind scenario as well as with the MHD wind scenario.  We also find, using both continuum and PSF subtraction, that [Ar\,II] traces a more compact emission. Having the spatial information of not just one but two possible wind tracing emission lines can put a better constraint on the wind mass loss rate and the wind launch radii which in turn can provide insight into whether the wind is thermally driven or MHD driven. 

A simple estimate of the wind mass loss rate ($\dot M_W$) can be obtained by assuming an EUV-driven fully-ionized wind. It can be written as an integral of mass density $\times$ wind velocity $\times$ cross-section area. To get the mass density, we need the electron density, which can be obtained from Equation 4 of \cite{Font2004}.

\begin{equation}
    n_o = n_g \times \left(\frac{r}{r_g}\right)^{-3/2}
\label{eq: 2}
\end{equation}

where, $r_g$ is the gravitational radius, and $n_g$ is the base electron density at $r_g$ and can be calculated using Equations 12 and 13 of \cite{Font2004}, respectively.

\begin{equation}
    r_g \approx 1.3 \times 10^{14} \left(\frac{M_\star}{1~M_\odot}\right) cm
\label{eq: 3}
\end{equation}

\begin{equation}
    n_g \approx 4 \times 10^{4} \left(\frac{M_\star}{1 M_\odot}\right)^{-3/2} \left(\frac{\Phi_{EUV}}{10^{41}~s^{-1}}\right)^{1/2} cm^{-3} 
\label{eq: 4}
\end{equation}

where M$_\star$ is the stellar mass and $\Phi_{\mathrm{EUV}}$ is the stellar EUV flux. For T\,Cha, we use stellar mass of 1.5 M$_\odot$ \citep{Olofsson2011}, and we have the EUV flux upper limit from \cite{Pascucci2014} calculated using cm free-free emission as 4.1 $\times$ 10$^{41}$~s$^{-1}$. Substituting these values in Equations \ref{eq: 3} and \ref{eq: 4} gives $r_g$ $\sim$ 13 au and the electron density there $\sim$ 4 $\times$ 10$^{4}$ cm$^{-3}$, respectively. Substituting these values in Equation \ref{eq: 2} provides the base electron density only as a function of radius. Using this, we can write the total mass loss rate as

\begin{equation}
    \dot M_W = \int_{0.1r_g}^{r_{max}} \mu m_H n_o \times v_w \times 4 \pi r dr
\label{eq: 5}
\end{equation}

where $\mu$ is the mean molecular weight (1.35), m$_H$ is the mass of a hydrogen atom, n$_o$ is substituted from Equation \ref{eq: 2}, v$_w$ is the speed of the wind leaving the disk, 0.1$r_g$ is the wind launch radii \citep[e.g.,][]{Alexander2008}, and r$_{max}$ is the maximum radial extent of the observed [Ne\,II] emission. The blue shift observed in the high spectral resolution [Ne\,II] spectrum, after accounting for inclination,  can be taken as the wind speed -- for T\,Cha, this is $\sim$ 12 km~s$^{-1}$ \citep[4 km~s$^{-1}$ blue-shift and 70~$\deg$ inclination,][respectively]{Pascucci2009,Huelamo2015}. We can take r$_{max}$ as the quadrature subtraction of [Ne\,II] HWHM and the PSF HWHM in the vertical ($\sim$ disk) direction $\sim$ 6.5 au (see Section \ref{sec:psfsub}). Integrating Equation \ref{eq: 5} and substituting the above values provides a wind mass loss rate of $\sim$ 6.5 $\times$ 10$^{-10}$ M$_{\odot}~\mathrm{yr}^{-1}$. Note that this estimate has an uncertainty of at least a factor of 2, propagated by the uncertainty in the stellar EUV flux \citep{Pascucci2014}. If the wind is indeed EUV-driven, the above estimate will act as an upper limit since we only have the upper limit for EUV flux. At the same time, if the wind is not fully ionized, the mass loss rate will be higher than the above estimate. Detailed modeling is carried out by Sellek et al. (submitted) using the line luminosities, line ratios as well as spatial extent of the lines detected in this work to estimate the wind mass loss rate as well as the wind launch radii.

\subsection{Location and nature of the spatially extended PAH features}
\label{sec:PAH}

PAHs are constituted of multiple interconnected Benzene rings and are known to play a major role in the chemistry of the disk atmosphere \citep[e.g.,][]{Anderson2017} as they influence the ionization state of the gas \citep{Thi2019}. Their ability to efficiently convert incident far-UV (FUV) radiation into thermal energy due to a high photoelectric yield \citep{Bakes1994} significantly impacts atmospheric mass loss rates \citep[e.g.,][]{Gorti2009b}. Since PAHs have high opacity at UV wavelengths, their radial distribution provides important information on the UV radiation field in the inner and outer disk \citep[e.g.,][]{Geers2007}. 

\begin{figure*}
    \centering
    \includegraphics[width=\textwidth]{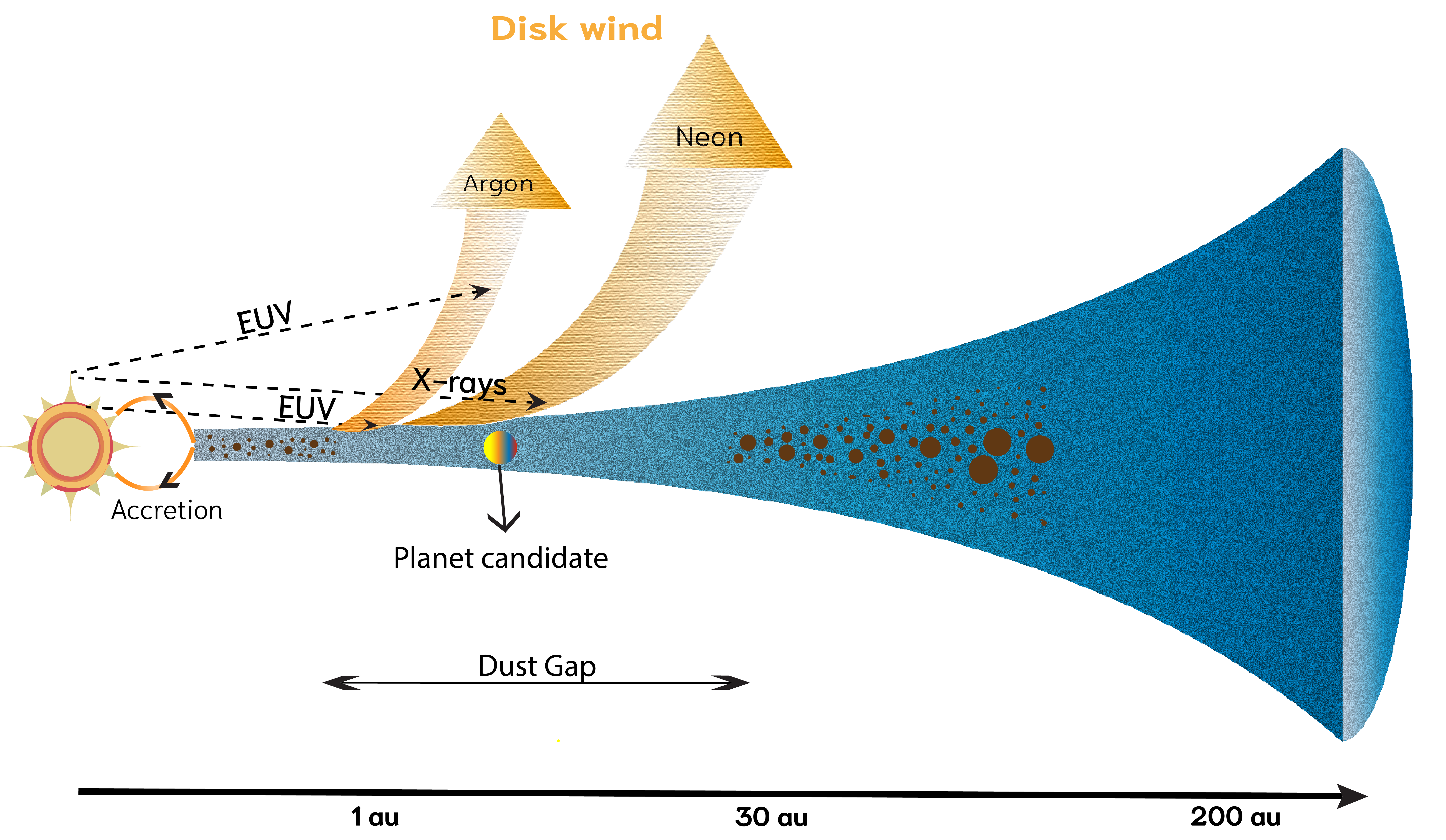}
    \caption{Sketch for T\,Cha highlighting the Argon and Neon ionization mechanisms as well as the relative spatial extent of the disk wind emission traced by the two species. EUV ionizes Argon, but it is blocked by the wind component traced by Argon. However, X-rays can penetrate this component and ionize Neon further out in the disk; see the more evolved disk/wind systems in \cite{Pascucci2020}. The location of the planet candidate is from \cite{Huelamo2011} and the dust gap is from \cite{Hendler2018}.}
    \label{fig:illustration}
\end{figure*}

We found that the PAH features in T\,Cha are more extended than the nearby continuum (Section \ref{sec:extended_wind}). They are found to be extended in the disk direction (vertical cut) instead of the wind direction (horizontal cut), suggesting that they are not being carried away in the wind but could still be transported by the stellar radiation along the disk surface. We also cannot exclude the possibility of the PAHs originating at the base of the wind. \cite{Geers2007} detected the 3.3 \micron{} and 11.2 \micron{} PAH features using the ISAAC L-band and VISIR N-band spectra of T\,Cha. Since both emissions were spatially unresolved, they provided an upper limit on the radial extent of these PAH emissions of 35\,au and 20\,au, respectively. \cite{Olofsson2013} modeled the dust disk of T\,Cha including the PAH emission, and compared model results to observations carried out with VLTI/MIDI and NACO/SAM. They suggested that PAH molecules cannot be located at the inner edge of the outer disk ($\sim12$\,au) as they would be too efficient at backward scattering (which is not observed) and might have been transported outward by radiation pressure or stellar wind. Although they could not provide an estimate on how radially extended the PAHs are, the fact that we found PAH emission more extended ($\sim$37 au along the disk) than the continuum is consistent with their argument.

Particularly for transition disks, \cite{Maaskant2014} found strong correlations between the relative integrated fluxes of the 6.2 \micron{} and the 11.2 \micron{} PAH features and their spatial extent. To infer this extent, they used VLT/VISIR and compared the FWHM of the spatially resolved PAH emission with the N-band continuum. They found that in case of stronger 6.2 \micron{} than 11.2 \micron{} flux (I$_{6.2}$/I$_{11.2}$ $\sim$ 2-4), the PAH emission is dominated by the inner regions including the dust gap. For similar strengths of the two features (I$_{6.2}$/I$_{11.2}$ $\sim$ 1-2), most PAH emission arises in the outer disk with some possible emission in the dust cavity. And finally, for stronger 11.2 \micron{} feature than 6.2 \micron{} (I$_{6.2}$/I$_{11.2}$ $<$ 1), the PAHs are more extended outwards and dominate only in the outer disk. \cite{Maaskant2014} modeled the PAH emission from four of these disks and found that having ionized PAHs in the dust gap increases the I$_{6.2}$/I$_{11.2}$ ratio and that a value $\sim$1 can be explained by neutral PAHs. For T\,Cha, we find I$_{6.2}$/I$_{11.2}$ $\sim$ 0.9 (see Table \ref{tab:intensities}) suggesting similar strengths for the two features. Therefore, PAHs being spatially extended in the MIRI MRS cubes is consistent with the trend observed by \cite{Maaskant2014}. The ratio also means that the PAHs in the disk of T\,Cha are predominantly neutral.

\cite{Geers2006} analysed the \textit{Spitzer}-IRS detected 11.2 \micron{} PAH feature towards T\,Cha. They noticed that this feature is both broader and red-shifted compared to a typical ISM PAH feature which peaks in the wavelength range of 11.2-11.24 \micron{} and has a width of $\sim$ 0.17 \micron. With the better resolution of MIRI MRS compared to Spitzer IRS, we find that the 11.2 \micron{} feature in the JWST spectra peaks at $\sim$11.29 \micron{} and a Gaussian fit gives a width of $\sim$0.25 \micron. Due to the lack of crystalline silicate features, the possibility of the 11.2 \micron{} crystalline silicate feature contributing to the broadening was eliminated, and it was rather shown that a model including a combination of fundamental and hot bands of PAH transition could explain the feature \citep{Geers2006}. Recently, \cite{Mackie2022} showed that the anharmonic interactions between different rotational modes of PAHs shift the emission spectra to higher wavelengths and broaden by the factor seen in several observations \citep[e.g. Orion Bar,][]{Peeters2002,Diedenhoven2004}. This reasoning was also strongly supported by the recent MIRI MRS observations of PAHs in the Orion Bar \citep{Chown2023}.

\section{Summary and Conclusions} \label{sec:sum&con}

Using JWST MIRI MRS, we observed T\,Cha, a G8 spectral type star surrounded by a disk with a dust gap ($\sim$20 au). It is known to have a small blue shift ($\sim$3-6 km~s$^{-1}$) in its [Ne\,II] line at 12.8 \micron{} indicating a slow wind that is at least partly ionized. We retrieved the IFU cube for T\,Cha over the complete MRS wavelength range of $\sim$5-28 \micron{}. We extracted the spectrum over 2.5 $\times$ PSF and searched for gas emission lines. We also performed spaxel-by-spaxel continuum subtraction in the image plane of T\,Cha to test if the strongest emission lines are spatially extended. On the continuum-subtracted line image, we performed a 2-D Gaussian fit to determine the emission size and compare it with the extension of the standard star/PSF. Finally, we performed PSF subtraction on these continuum-subtracted images and retrieved the PSF subtracted maps of [Ne\,II] and [Ar\,II]. The main conclusions from this work are listed below:
\begin{itemize}
    \item In addition to [Ne\,II] which was previously detected and known to trace a wind, we report for the first time detections of [Ar\,II] at 6.98 \micron{}, [Ar\,III] at 8.99 \micron{}, and [Ne\,III] at 15.55 \micron{}.  We also find individual PAH features (6.02 \micron{}, 8.22 \micron{}) as well as PAH complexes (6.2 \micron{}, 7.7 \micron{}, 8.6 \micron{}, 11.3 \micron{}, 12 \micron{}, and 12.7 \micron{}) of which only the 11.3 \micron{} complex was previously detected for T\,Cha. Additionally, we also detect the H$_2$v0-0 S(3) line in the spectrum of T\,Cha.  
    \item We find that the MIRI [Ne\,II] flux is 50\% higher than the \textit{Spitzer} flux obtained in 2006. The amplitude of [Ne\,II] variability (50\%) is smaller than that observed for [O\,I] ($\sim$ 230\%), and it is unclear whether variabilities in these two lines are correlated to each other or not. Future simultaneous observations of both lines can help us understand the evolution of these wind diagnostic lines better. 
    \item  By using the [Ne\,II]/[Ne\,III] and [Ar\,II]/[Ar\,III] line ratios and ionization model predictions from \cite{Hollenbach2009}, it is most likely that Argon is being ionized by EUV while Neon is being ionized by X-rays in T\,Cha. However, Soft X-ray ionization of Argon cannot be ruled out just yet.
    \item  We find that [Ne\,II] and [Ne\,III] trace extended emission beyond the PSF while [Ar\,II] traces a relatively compact emission in T\,Cha.
    \item  We also find that the PAH emission is spatially more extended than the continuum, suggesting that the emission is dominated in the outer disk surface. Based on the relative strengths of the 6.2 \micron{} and 11.2 \micron{} features, we find that the PAH emission is mostly neutral.
\end{itemize}

T\,Cha is the first protoplanetary disk for which four forbidden noble gas lines are detected together, as well as the first detection of [Ar\,III]. We suggest that the higher MIRI [Ne\,II] flux could result from increased wind mass loss rate, more heating, or higher fractional ionization caused by higher radiation received by the outer disk. The more compact emission traced by [Ar\,II] can act as a screen for photons ionizing Ne and Ne$^+$ atoms further out in the disk, see Figure~\ref{fig:illustration}. This is in agreement with Argon being ionized by EUV and Ne and Ne$^+$ being ionized by X-rays. At the same time, the spatial extension of the [Ne\,II] and [Ne\,III] emission is consistent with both a PE wind and an MHD wind. However, T\,Cha is a weak accretor that shows no evidence of a jet, has a relatively weak but broad (FWHM$\sim$47 km~s$^{-1}$) [O\,I] 6300\,\AA\ line centered at the stellar velocity, and a mid-infrared spectrum dominated by ionic lines. Putting this information together suggests that T\,Cha might represent the more evolved wind/disk stage shown in the bottom panel of Figure 9 of \cite{Pascucci2020}. The [O\,I] would be tracing a very close-in but tenuous MHD wind, while ionized Argon and Neon may be tracing a PE wind further out, with Argon tracing a more compact wind structure than Neon. This picture is in agreement with results from the radiative transfer photoionization models of disk winds in T\,Cha presented in Sellek et al. (submitted). Detailed modeling combining MHD and PE winds is necessary to further test this evolutionary scenario put forward for T\,Cha. Additional MIRI MRS observations of low accreting stars with gapped disks will be critical to determine the prevalence of spatially extended winds traced by ions, as observed in T\,Cha.

\section{Acknowledgements}
This work is based on observations made with the NASA/ESA/CSA James Webb Space Telescope. The data were obtained from the Mikulski Archive for Space Telescopes at the Space Telescope Science Institute, which is operated by the Association of Universities for Research in Astronomy, Inc., under NASA contract NAS 5-03127 for JWST. The observations are associated with JWST GO Cycle 1 program ID 2260. The JWST data used in this paper can be found in MAST: \href{http://dx.doi.org/10.17909/dhmh-fx64}{10.17909/dhmh-fx64}. This work was funded by the NASA/STScI GO grant JWST-GO-02260.001. A.D.S. was supported by the European Union’s Horizon 2020 research and innovation programme under the Marie Sklodowska-Curie grant agreement number 823823 (DUSTBUSTERS). G.B. has received funding from the European Research Council (ERC) under the European Union’s Horizon 2020 Framework Programme (grant agreement no. 853022, PEVAP). N.S.B. Thanks Arin Avsar for the useful discussion on PSF subtraction technique. N.S.B. acknowledges the quick response and support from the JWST MIRI support team at STScI. R.A. acknowledges funding from the Science \& Technology Facilities Council (STFC) through Consolidated Grant ST/W000857/1. A.D.S. acknowledges the support of a Science and Technology Facilities Council (STFC) Ph.D. studentship and funding from the European Research Council (ERC) under the European Union’s Horizon 2020 research and innovation programme (grant agreement No. 1010197S1 MOLDISK). This work is based in part on archival observations made with the \textit{Spitzer} Space Telescope, which was operated by the Jet Propulsion Laboratory, California Institute of Technology under a contract with NASA. This work has also made use of data from the European Space Agency (ESA) mission Gaia \url{https://www.cosmos.esa.int/gaia}, processed by the Gaia Data Processing and Analysis Consortium (DPAC, \url{https://www.cosmos.esa.int/web/gaia/dpac/consortium}). Funding for the DPAC has been provided by national institutions, in particular, the institutions participating in the Gaia Multilateral Agreement.

\vspace{5mm}
\facilities{James Webb Space Telescope, Spitzer Space Telescope, Gaia}

\software{Astropy \citep{astropy2013,astropy2018,astropy2022}, CASA \citep{McMullin2007}, JWST \citep{Bushouse2023}, Matplotlib \citep{Hunter2007}, NumPy \citep{vanderwalt2011}, Photutils \citep{Bradley2022}, SciPy \citep{Virtanen2020}}

\bibliography{citations.bib}{}
\bibliographystyle{aasjournal}

\appendix

\section{Gaussian fit to lines}
\label{sec:gauss_fit}

\begin{figure}
    \centering
    \includegraphics[width=\textwidth]{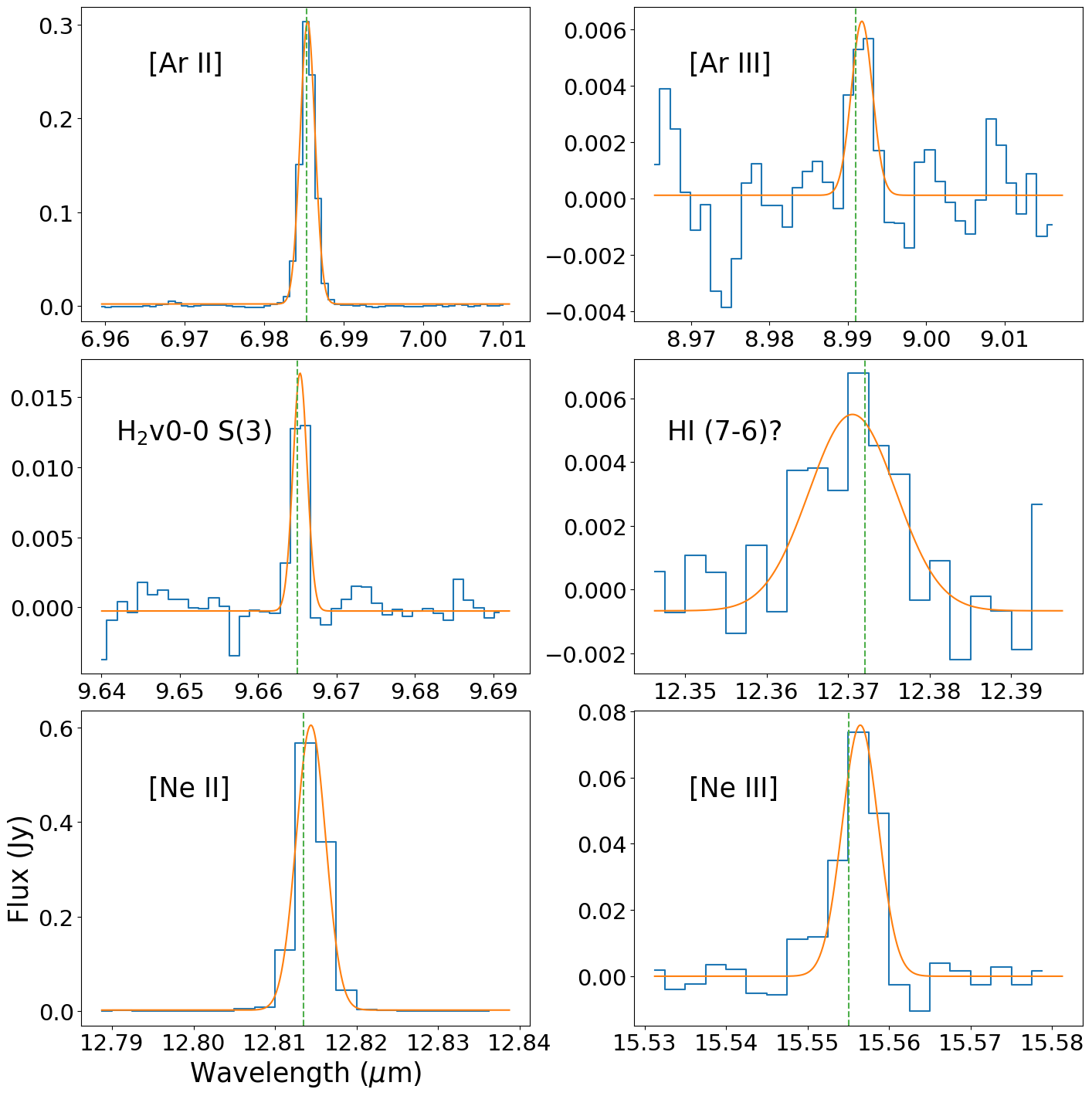}
    \caption{The figure shows JWST MIRI MRS spectral profiles for the lines listed in Table \ref{tab:intensities} in blue and the performed Gaussian fit in orange. The green dashed line indicates the rest wavelength of the transition. Each subplot has the same length of the wavelength range ($\sim$0.055 \micron{}) plotted on the x-axis with the lines peaking at the center.}
    \label{fig:gauss}
\end{figure}

We fit Gaussian profiles to the lines listed in Table \ref{tab:intensities} and plot them in Figure \ref{fig:gauss}. [Ar\,III] has a 4 $\sigma$ detection, whereas other lines, except HI (7-6), have $\geq$ 10 $\sigma$ detections. Since each subplot has an equal x-axis length, the width of the Gaussian profile increases with wavelength as the spectral resolution degrades with increasing wavelength. This can be seen in Figure \ref{fig:gauss} where the width is smallest for [Ar\,II] 6.98 \micron{} and largest for [Ne\,III] 15.55 \micron{}, as expected. For HI (7-6), while there is an emission with a peak amplitude of $\sim$ 3$\sigma$ at its position, the width of the fitted line is $\sim$ 3 times the spectral resolution of MIRI MRS ($\sim$ 100 km~s$^{-1}$) at that wavelength, which is rather unrealistic. Hence, we report HI (7-6) as a non-detection and provide an upper limit. To cross-check the [Ar\,III] detection, we compare spectra from different reductions by switching \texttt{residual\_fringe} step on and off. We also extract spectra using different apertures (1.5 $\times$ FWHM - 3 $\times$ FWHM, in four steps) and calculate the [Ar\,III] line flux in each and find that the line is always present and its flux varies within the uncertainties.

\section{PSF Subtraction using HD167060}
\label{sec:psf2sub}

\begin{figure}
    \centering
    \includegraphics[width=\textwidth]{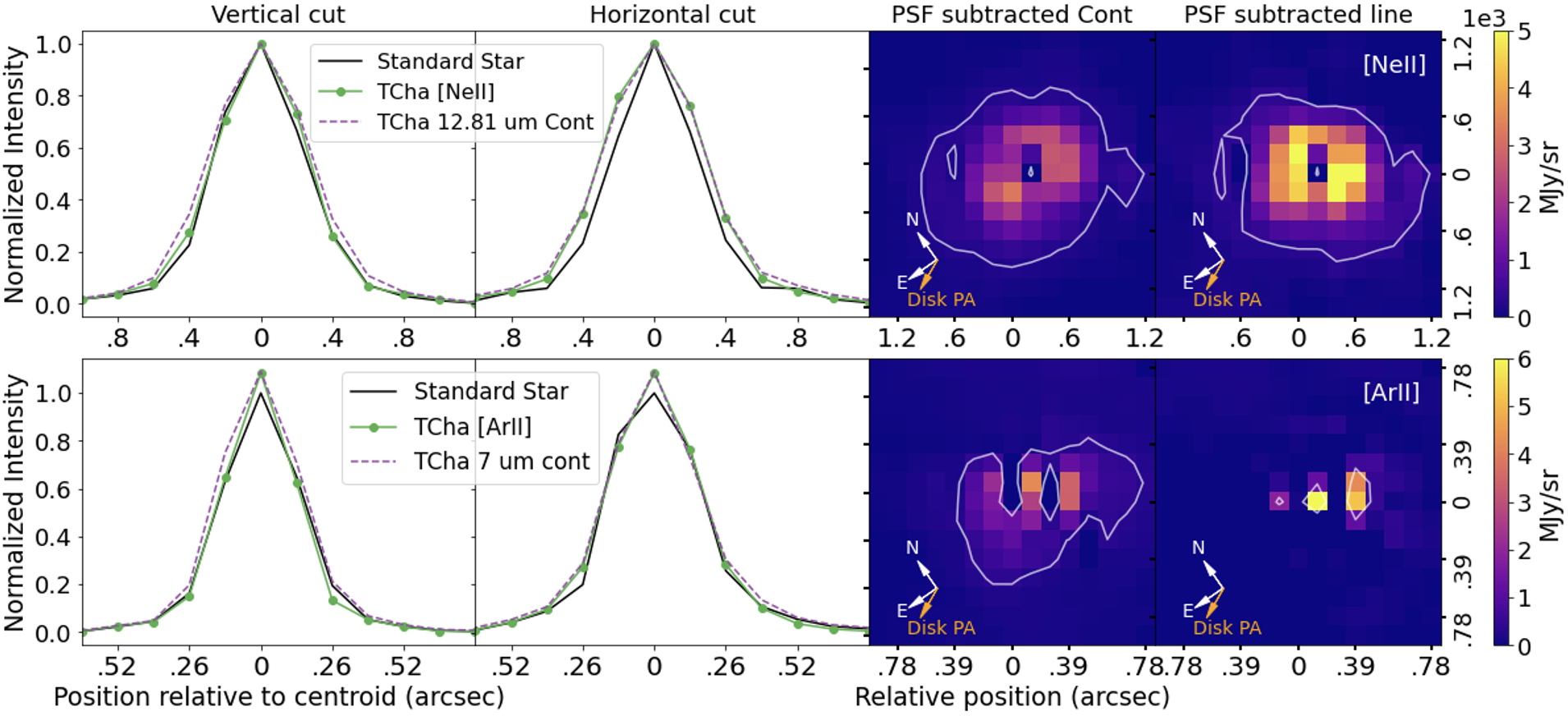}
    \caption{The left two panels compare the radial profiles of continuum and continuum-subtracted lines with the PSF (standard star HD167060) in the horizontal and vertical directions. The two panels on the right show the PSF-subtracted images of the continuum at 12.81 and 7 \micron{} and [Ne\,II] and [Ar\,II] lines. The white contours in the images follow 5$\sigma$ emission. }
    \label{fig:psf2sub}
\end{figure}

This section shows PSF subtraction results obtained by following the same method followed in Section \ref{sec:psfsub}, but by using the standard star HD167060 as the PSF instead of HD37962. The results are plotted in Figure \ref{fig:psf2sub}. At 12.81 \micron{} both the continuum and [Ne\,II] show emission beyond the PSF and have different directions of extension, consistent with the results described in the main text. Similarly, at 7 \micron{}, the continuum is very slightly resolved and [Ar\,II] shows very compact unresolved emission. A slight difference in the emission structure at 7 \micron{} arises due to the PSF's left shoulder showing extra emission. This could either be due to centroid fitting error or an interpolation error during image shifting. In either case, our results remain unaffected, and we are able to strongly demonstrate the independence of our results to the small PSF variations. 

\section{Extended PAH emission}
\label{sec:extended_pah}

\begin{figure}
    \centering
    \includegraphics[width=\textwidth]{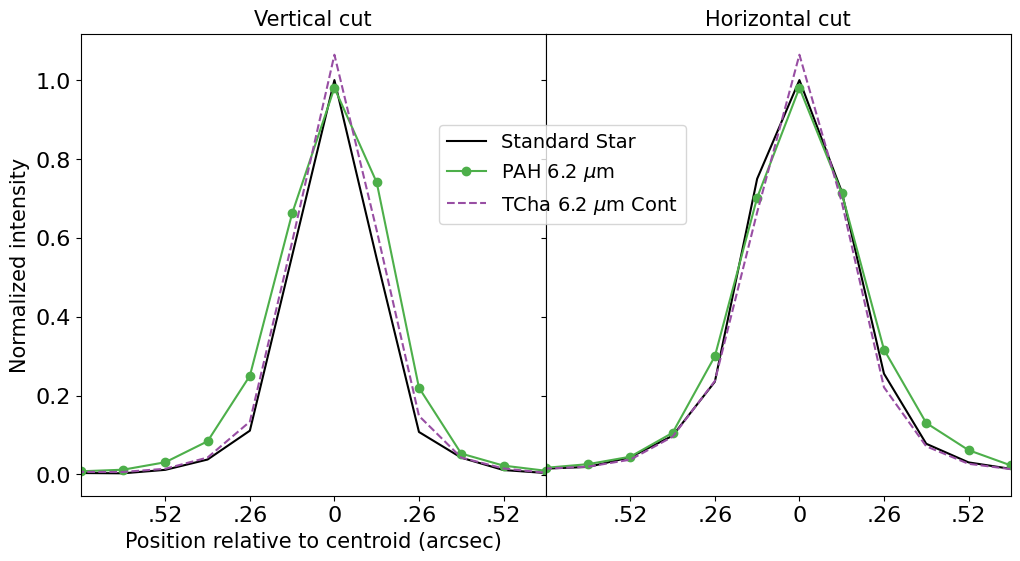}
    \caption{The two panels compare the radial profiles of continuum and continuum-subtracted PAH emission with the PSF/Standard star in the horizontal and vertical directions.}
    \label{fig:pah_psf}
\end{figure}

This section presents PSF comparison with the continuum-subtracted PAH emission feature at 6.2 \micron{} by following the procedure detailed in Section \ref{sec:psfsub}. Here, we use the standard star HD37962 as the PSF, as done for all the results in the main text. We show the vertical and horizontal profiles of the PSF, continuum, and continuum-subtracted PAH in Figure \ref{fig:pah_psf}. It can be easily seen that the PAH emission is extended beyond the PSF and the continuum in the vertical cut/profile. In terms of the direction of extension, this is consistent with the results shown in Section \ref{sec:psfsub} and \ref{sec:psf2sub}, where the continuum, when extended, is extended in the vertical profiles and wind in the horizontal profile. Since PAH is supposed to be tracing the disk surface, it should be extended mainly in the vertical profile, which is what we see in Figure \ref{fig:pah_psf}.

\end{document}